# A new model for virtual machine migration in virtualized cluster server based on Fuzzy Decision Making

M.Tarighi, S.A.Motamedi, S.Sharifian

**Abstract**—In this paper, we show that performance of the virtualized cluster servers could be improved through intelligent decision over migration time of Virtual Machines across heterogeneous physical nodes of a cluster server. The cluster serves a variety range of services from Web Service to File Service. Some of them are CPU-Intensive while others are RAM-Intensive and so on. Virtualization has many advantages such as less hardware cost, cooling cost, more manageability. One of the key benefits is better load balancing by using of VM migration between hosts. To migrate, we must know which virtual machine needs to be migrated and when this relocation has to be done and, moreover, which host must be destined. To relocate VMs from overloaded servers to underloaded ones, we need to sort nodes from the highest volume to the lowest. There are some models to finding the most overloaded node, but they have some shortcomings. The focus of this paper is to present a new method to migrate VMs between cluster nodes using TOPSIS algorithm - one of the most efficient Multi Criteria Decision Making techniques- to make more effective decision over whole active servers of the Cluster and find the most loaded serversTo evaluate the performance improvement resulted from this model, we used cluster Response time and Unbalanced Factor.

——————————— ◆ ———————————

## 1 INTRODUCTION

Operating system virtualization has attracted consider-able interest in recent years; particularly from the data center and cluster computing communities [6]. Data centers have become popular in a variety of domains such as web hosting, enterprise systems, and e-commerce sites [50]. Server resources in a data center are multiplexed across multiple applications--each server runs one or more applications. Further, each application sees dynamic workload fluctuations caused by incremental growth, time-of-day effects, and flash crowds [1].

Process migration, a hot topic in systems research during the 1980s [10, 11, 12, 13, 14], has seen very little use for real-world applications. Milojicic et al [65] give a thorough survey of possible reasons for this, including the problem of the residual dependencies that a migrated process retains on the machine from which it migrated. These are undesirable because the original machine must remain available, and because they usually negatively impact the performance of migrated processes.

Another Applicable approach for reducing management complexity is to employ virtualization. In this approach, applications run on virtual servers that are constructed using virtual machines, and one or more virtual servers are mapped onto each physical server in the system. Virtualization of data center resources provides numerous benefits. It enables application isolation since malicious or greedy applications can not impact other applications co-located on the same physical server. It enables server consolidation and provides better multiplexing of data center resources across applications. Perhaps the biggest advantage of employing virtualization is the ability to flexibly remap physical resources to virtual servers in order to handle workload dynamics. Migration is transparent to the applications and all modern virtual machines support this capability [6,20].

A workload increase can be handled by increasing the resources allocated to a virtual server, if idle resources are available on the physical server, or by migrating the virtual server to a less loaded physical server.

The Xen migration work [6] showed that virtual machine migration can enable robust and highly responsive provisioning in data centers. What is missing is a convincing validation and algorithms to effect migration,

The idea of process migration was first investigated in the 80's [27]. Support for migrating groups of processes across

———————————————

- *M.Tarighi is with the Electrical & Electronics Engineering Department, University of AMIR KABIR, High Performance Computing Lab., Tehran, Iran.*
- *S.A.Motamedi is with the Electrical & Electronics Engineering Department, University of AMIR KABIR, High Performance Computing Lab., Tehran, Iran.*
- *S.Sharifian is with the Electrical & Electronics Engineering Department, University of AMIR KABIR, High Performance Computing Lab., Tehran, Iran.*
- 

OSes was presented in [16], but applications had to be suspended and it did not address the problem of maintaining open network connections. Virtualization support for commodity operating systems in [7] led towards techniques for virtual machine migration over long time spans, suitable for WAN migration [24]. More recently, Xen [6] and VM-Ware [20] have implemented ``live'' migration of VMs that involve extremely short downtimes ranging from tens of milliseconds to a second. VM migration has been used for dynamic resource allocation in Grid environments [23,26 ,8]. A system employing automated VM migrations for scientific nano-technology workloads on federated grid environments was investigated in [23]. The Shirako system provides infrastructure for leasing resources within a federated cluster environment and was extended to use virtual machines for more flexible resource allocation in [8]. Shirako uses migrations to enable dynamic placement decisions in response to resource broker and cluster provider policies. In contrast, we focus on data center environments with stringent SLA requirements that necessitate highly responsive migration algorithms for online load balancing. VMware's Distributed Resource Scheduler [29] uses migration to perform automated load balancing in response to CPU and memory pressure. DRS uses a userspace application to monitor memory usage similar to Sandpiper's gray box monitor, but unlike Sandpiper, it cannot utilize application logs to respond directly to potential SLA violations or to improve placement decisions.

Dedicated hosting is a category of dynamic provisioning in which each physical machine runs at most one application and workload increases are handled by spawning a new replica of the application on idle servers. Physical server granularity provisioning has been investigated in [1,22]. Techniques for modeling and provisioning multi-tier Web services by allocating physical machines to each tier are presented in [28]. Although dedicated hosting provides complete isolation, the cost is reduced responsiveness - without virtualization, moving from one physical machine to another takes on the order of several minutes [28] making it unsuitable for handling flash crowds. Our current implementation does not replicate virtual machines, implicitly assuming that PMs are sufficiently provisioned.

Shared hosting is the second variety of dynamic provisioning, and allows a single physical machine to be shared across multiple services. Various economic and resource models to allocate shared resources have been presented in [5]. Mechanisms to partition and share resources across services include [2,5]. A dynamic provisioning algorithm to allocate CPU shares to VMs on a single physical machine (as opposed to a cluster) was presented and evaluated through simulations in [19]. In comparison to the above systems, our work assumes a shared hosting platform and uses VMs to partition CPU, memory, and network resources, but additionally leverages VM migration to meet SLA objectives.

Estimating the resources needed to meet an appliction's SLA requires a model that inspects the request arrival rates for the application and infers its CPU, memory, and network bandwidth needs. Developing such models is not the focus of this work and has been addressed by several previous efforts such as [17,1].

Our work is to present a new model to migrate VMs between cluster nodes using TOPSIS algorithm to make decision over whole active servers of the data center and find the most loaded server.We have implemented our techniques using the Xen virtual machine [3]. We conduct a detailed experimental evaluation on cluster servers using a mix of CPU-, network- and memory-intensive applications. Results show that our techniqe imposes some overheads rather than Sandpiper but it makes better migration decisions when hotspot occurs.

In order to introduce the suggested fuzzy decision making model based on TOPSIS, firstly, disadvantages of the existing method of ranking the servers will be presented. Then, the basic concepts of the technique for order preference by similarity to an ideal solution1 and the Fuzzy TOPSIS are mentioned. Moreover, the fuzzy decision making2 software tool is introduced. An application of the FDM software tool is carried out through case studies. The rest of this paper is structured as follows. Section 2 presents some existing methods of finding most loaded servers and their limitations, and Sections 3-6 present our designed algorithm. Section 7 presents our implementation and evaluation. Finally, Sections 8 and 9 present background and related Work and our conclusions, respectively.

## 2 BACKGROUND AND RELATED WORK

Detecting workload hotspots and initiating a migration is currently handled manually. Manually-initiated migration lacks the ability to respond quickly to sudden workload changes.

To address this challenge, studies automated strategies for virtual machine migration in large data centers. The proposed techniques automate the tasks of monitoring system resource usage, hotspot detection, determining a new mapping and initiating the necessary migrations. Migration is further complicated by the need to consider multiple resources-CPU, network,and memory-for each application and physical server [67].

---

[1] TOPSIS
[2] FDM

## 2.1 Existing methods of finding most loaded servers and their limitations

Sandpiper implements a hotspot detection algorithm that determines when to migrate virtual machines, and a hotspot mitigation algorithm that determines what and where to migrate [67]. Upon hotspot detection, Sandpiper's migration manager is invoked for hotspot mitigation. The migration manager employs provisioning techniques to determine the resource needs of overloaded VMs and uses a greedy algorithm to determine a sequence of moves or swaps to migrate overloaded VMs to underloaded servers. The migration manager determines which VMs need to be migrated.

The basic idea is to move load from the most overloaded servers to the least-overloaded servers, while attempting to minimize data copying incurred during migration. Since a VM or a server can be overloaded along one or more of three dimensions– CPU, network and memory–it defined a new metric that captures the combined CPU-network-memory load of a virtual and physical server. The volume of a physical or virtual server is defined as the product of its CPU, network and memory loads:

$$vol = \frac{1}{1-cpu} * \frac{1}{1-net} * \frac{1}{1-mem} \tag{1}$$

Where CPU, net and mem are the corresponding utilizations of that resource for the virtual or physical server. The higher the utilization of a resource, the greater the volume; if multiple resources are heavily utilized the above product results in a correspondingly higher volume. Unfotunately, In such numerical ranking methods, the influence of each parameter is verified separately and the mutual effects of parameters are ignored.Also in these , all the criteria are assumed to have equal weights in decision making, but considering the status of each parameter makes such an assumption unrealistic.Fore example, when the virtual machines are web server and because the Web servers are cpu-intensive load, the propoabilty of CPU saturation is more than RAM saturation. In the other word, the weight of CPU and RAM influence are not equal.

VMware has added OS migration support, dubbed VMotion, to their VirtualCenter management software [69]. VMware Distributed Resource Scheduler improves resource allocation across all hosts and resource pools in a cluster. When a cluster is enabled for DRS, VirtualCenter continuously monitors the distribution of CPU and memory resource usage for all hosts and virtual machines in that cluster. DRS compares these metrics to ideal resource utilization—that is, the virtual machines' entitlements. These entitlements are determined based on the resource policies of virtual machines in the cluster and their current demands. VirtualCenter uses this analysis to perform initial placement of virtual machines, virtual machine migration for load balancing, enforcement of rules and policies [68]. As the number of virtual machines and host increases, overhead imposed by balancing migration increases too. Moreover; the cluster which is setup to run DRS consists of a homogenous configuration of hosts. This assumption simplify the decision algorithm and reduse the overhead. Another negative point of DRS is that DRS does not make virtual machine placement decisions based on their usage of I/O resources. VMware's DRS will use algorithms based on CPU and memory to decide how to balance the hosts. If some I/O-intensive workloads share a single host, they might saturate the host's I/O capacity, leading to performance degradation.In the other word, parameters which algorithm of sorting physical node considered are restricted to CPU and RAM status. In addition,As this is commercial software and strictly disallows the publication of third-party benchmarks, we are only able to infer its behavior from VMware's own publications. These limitations make a thorough technical comparison impossible. However, based on the VirtualCenter User's Manual [66], their approach is generally similar to XEN live migration [6] and would expect it to perform to a similar standard.

## 3 METHODOLOGY AND THE PROPOSED MODEL

Considering limitations and disadvantages of existing method of calculating criticality of cluster servers, efforts have been made in order to develop decision making models for sorting physical nodes.

The existing decision making models for server selection are useful but have restricted applications. These methods cannot deal with decision maker ambiguities, uncertainties and vagueness, which cannot be handled by crisp values. Having to use crisp values is one of the important problematic points in their process. In this article, the concept of the approach used for sorting problem is based on the fuzzy technique for order preference by similarity to ideal solution (Fuzzy TOPSIS). This is because four advantages are addressed: (1) a sound logic that represents the rationale of human choice, (2) a scalar value that accounts for both the best and worst alternatives simultaneously, (3) a simple computation process that can be easily programmed, and (4) the performance measures of all alternatives on attributes can be visualized. These advantages make TOPSIS a major decision making technique as compared with other related techniques such as AHP[26, 27]. The disadvantages of the AHP technique are that it focuses mainly on the decision maker who has to make many pair-wise comparisons to reach a decision, while possibly using subjective preferences. Furthermore, an important disadvantage of the AHP method is the artificial limitation of the use of the nine-point scale. For instance, if Alternative A is five times more important than Alternative B, which in turn is five times more important than Alternative C, a serious evaluation problem arises. The Saaty method[32] cannot cope with the fact that Alternative A is twenty five times more important than Alternative C[33]. The methodology is useful only when the decision making framework has a unidirectional hierarchical relationship among decision levels. Moreover, AHP is not practically usable if the number of alternatives and criteria are large since the repetitive assessments may cause fatigue in the decision maker[34,35].

Our algorithm have a hierarchial structure. A supervisor program is monitoring status of all nodes based on usage statistics gathered from various physical servers. Workload of each node varies over time and may cross the threshold. So, to avoid server saturation and service down time the program check the physical host to find overloaded ones. Upon hotspot detection in a node. The level two program will be run. After finding the appropriate VM from the most loaded node identified in level 1 program, The third program which is responsible to migrate VMs operates automatically.All of these programs encapsulated in a software package mplemented in DOM 0 of the control node.

We assumed that the cluster is heterogeneous; therefore, lots of parameters like CPU clock speed, RAM capacity, RAM usage, NET usage, operating temperature… of hosts and virtual machines are used to make decision.

Recent methods for decision making processes have enabled decision-makers to decide more quickly, easily and sensitively[30]. Therefore; we can apply them to constract a new model to choose the best virtual machine to be moved from overloaded to underloaded servers to better load balancing. The deterministic explanation in numerical ranking methods can be mentioned as the most important disadvantage, because in the method offered by Sandpiper, parameters are explained by crisp values. But some parameters can be linguistic statements. For example, for temperature of each cluster node, linguistic terms are better to be used. Therfore; we divided Parameters affecting on node ranking into three classes of crisp3, linguistic and fuzzy parameters to add more flexibility in deciding.

### 3.1 TOPSIS AND FUZZY TOPSIS

TOPSIS is a popular approach to the MCDM method and has been widely used in the literature (Abo-Sinna and Amer [37]; Agrawal et al.[38]; Cheng et al.[39]; Deng et al.[40]; Feng and Wang[41, 42]; Hwang and Yoon[43]; Jee and Kang[44]; Kim et al.[45]; Lai et al.[46]; Liao[47]; Olson[48]; Opricovic and Tzeng[49]; Parkan and Wu[50,51]; Tong and Su[52]; Tzeng et al.[53]; Zanakis et al.[54]). The method has also been extended to deal with Fuzzy MCDM problems. For example, Chen [55] first converted a fuzzy MCDM problem into a crisp one via centroid defuzzification and then solved the nonfuzzy MCDM problem using that method. Chen and Tzeng [56] transformed a fuzzy MCDM problem into a nonfuzzy MCDM using a fuzzy integral. Instead of using distance, they employed the grey relation grade to define the relative closeness of each alternative. Chu [57, 58] and Chu and Lin [59] also changed a fuzzy MCDM problem into a crisp one and solved the crisp MCDM problem using the method. Differing from the others, they first derived the membership functions of all the weighted rankings in a weighted normalization decision matrix using interval arithmetics of fuzzy numbers and then defuzzified them into crisp values using the ranking method of mean of removals (Kaufmann and Gupta [60]).

Chen[55] extended the method to fuzzy group decision making situations by defining a crisp Euclidean distance between any two fuzzy numbers. Triantaphyllou and Lin [61] developed a fuzzy version of the method based on fuzzy arithmetic operations, which led to a fuzzy relative closeness for each alternative proposed by Wang and Elhag [62].

The TOPSIS method is a technique for order preference by similarity to ideal solution and proposed by Hwang and Yoon [43]. The ideal solution (also called the positive ideal solution) is a solution that maximizes the benefit criteria/attributes and minimizes the cost criteria/attributes, whereas the negative ideal solution (also called the anti-ideal solution) maximizes the cost criteria/attributes and minimizes the benefit criteria/attributes. The so-called benefit criteria/attributes are those for maximization, while the cost criteria/attributes are those for minimization. The best alternative is the one that is closest to the ideal solution and farthest from the negative ideal solut ion. Suppose a MCDM problem with m alternatives, $A1,...,Am$, and n decision criteria/attributes, $C1,...,Cn$. Each alternative is evaluated with respect to m criteria/attributes. All the values/ratings assigned to the alternatives with respect to each criterion form a decision matrix denoted by $X = (xij)$ nxm. Let $W = (w1,.., wn)$ be the relative weight vector for the criteria, satisfying $\sum ni=1wi = 1$, then the method can be summarized as follows[34]

---

3 deterministic

a) Calculate the decision matrix (D) as:

$$(2)$$

$$D = \begin{bmatrix} x_{11} & \cdots & x_{1n} \\ \vdots & \cdots & \vdots \\ x_{m1} & \cdots & x_{mn} \end{bmatrix}$$

b) Calculate the normalized decision matrix or R matrix. The normalized value $r_{ij}$ is calculated as:

$$(3)$$

$$R = \begin{bmatrix} r_{11} & \cdots & r_{1n} \\ \vdots & \cdots & \vdots \\ r_{m1} & \cdots & r_{mn} \end{bmatrix}$$

$$r_{ij} = \frac{r_{ij}}{\sqrt{\sum_{i=1}^{m} r_{ij}^2}} \qquad (4)$$

c) Calculate the criteria weighted matrix as:

$$(5)$$

$$W = \begin{bmatrix} w_1 & \cdots & 0 \\ \vdots & w_2 \cdots & \vdots \\ 0 & \cdots & w_n \end{bmatrix}$$

d) Calculate the weighted normalized decision matrix. The weighted normalized value $v_{ij}$ is calculated as:

$$(6)$$

$$v_{ij} = w_j r_{ij} = W \times R \quad j = 1,...,m, i = 1,...,n$$

Where wj is the weight of the jth criterion and $\sum_{i=1}^{n} wi = 1$

e) Determine the positive ideal and negative ideal solution, A+ and A− respectively.

$$(7)$$

$$A^+ = \{v_1^*,...,v_n^*\} = \left\{ \left( \max_j v_{ij} | i \in I \right), \left( \min_j v_{ij} | i \in J \right) \right\}$$

$$(8)$$

$$A^- = \{v_1^-,...,v_n^-\} = \left\{ \left( \min_j v_{ij} | i \in I \right), \left( \max_j v_{ij} | i \in J \right) \right\}$$

Where I is associated with benefit criteria, and J is associated with cost criteria.

f) Calculate the separation measures, using the ndimensional Euclidean distance. The distance of each alternative from the ideal solution is given as:

$$(9)$$

$$d_j^- = \sqrt{\sum_{i=1}^{n} \left( v_{ij} - v_i^- \right)^2} \quad j = 1,...,m$$

Similarly, the distance from the negative ideal solution is

given as:

$$(10)$$

$$d_j^- = \sqrt{\sum_{i=1}^{n} \left( v_{ij} - v_i^- \right)^2} \quad j = 1,...,m$$

g) Calculate the relative closeness to the ideal solution. The relative closeness of the alternative Aj with respect to A+ is defined as:

$$(11)$$

$$RC_i = \frac{d_i^-}{d_i^* + d_i^-}$$

Since d− J ≥ 0 and d +J ≥ 0, then clearly RCj ∈ [0,1] .

h) Rank the alternatives according to the relative closeness to the ideal solution: the higher RCj, the better alternative Aj. [62].

The fuzzy theory is a modern theory, which was proposed by Zadeh [63]. In classic logic, events have two values: to be or not to be, to exist or not to exist, black or white, and one or zero. But in fuzzy logic, in order to answer to events, a consistent spectrum is considered between 'to exist' and 'not to exist' and world phenomena are seen as gray—neither black nor white. The use of fuzzy theory allows us to incorporate unquantifiable information, incomplete information, non-obtainable information, and partial facts into the decision model. The fuzzy decision matrix (D~) and criteria weighted (W ~) can be concisely expressed in matrix format as:

$$(12)$$

$$\widetilde{D} = \begin{bmatrix} \widetilde{x}_{11} & \cdots & \widetilde{x}_{1n} \\ \vdots & \cdots & \vdots \\ \widetilde{x}_{m1} & \cdots & \widetilde{x}_{mn} \end{bmatrix}$$

$$(13)$$

$$\widetilde{W} = \begin{bmatrix} \widetilde{w}_1 & \cdots & 0 \\ \vdots & \cdots & \vdots \\ 0 & \cdots & \widetilde{w}_n \end{bmatrix}$$

where x~ij, i = (1,2,...,m), j = (1,2,...,n)and w ~ j, j = (1,2,...,n) are fuzzy numbers, x~ij = (aij, bij, cij) and w ~ j = (wj1, wj2, wj3). That x~ ij is the performance rating of the ith alternative, Ai, with respect to the jth criteria, Cj and w ~j represents the weight of the jth attribute, Cj. The normalized fuzzy decision matrix denoted by R~is shown as:

$$(14)$$

$$\widetilde{R} = \begin{bmatrix} \widetilde{r}_{11} & \cdots & \widetilde{r}_{1n} \\ \vdots & \cdots & \vdots \\ \widetilde{r}_{m1} & \cdots & \widetilde{r}_{mn} \end{bmatrix}$$

If x~ij = (aij, bij, cij), i = (1,2,...,m) and j = (1,2,...,n) are triangular fuzzy numbers, then the normalization process can be conducted by:[62]

$$(15)$$

$$\widetilde{r}_{ij} = \left( \frac{a_{ij}}{c_j^+}, \frac{b_{ij}}{c_j^+}, \frac{c_{ij}}{c_j^+} \right), \ i = 1,2,...,m; j \in \Omega_b$$

$$(16)$$

$$\widetilde{r}_{ij} = \left( \frac{a_j^-}{c_{ij}}, \frac{a_j^-}{b_{ij}}, \frac{a_j^-}{a_{ij}} \right), \ i = 1,2,...,m; j \in \Omega_c$$

Where $\Omega_b$ and $\Omega_c$ are the sets of benefit criteria and cost criteria, respectively and c+ j = max I cij, j ∈ $\Omega_b$ and a− j = min I aij, j ∈ $\Omega_c$. The weighted fuzzy normalized decision matrix is shown as:

$$(17)$$

$$\widetilde{v}_{ij} = \widetilde{w}_j \, \widetilde{r}_{ij} = \widetilde{W} \times \widetilde{R} \ \ j = 1,...,m, \ \ i = 1,...,n$$

The fuzzy positive-ideal (A+) and the fuzzy negativeideal (A−) solutions are shown as:

$$(18)$$

$$A^+ = (\widetilde{v}_1^+, \widetilde{v}_2^+, ..., \widetilde{v}_n^+) = \left\{ \max_1 v_{ij} \middle| \begin{matrix} i = 1,2,...,m \\ j = 1,2,...,n \end{matrix} \right\}$$

$$(19)$$

$$A^- = (\widetilde{v}_1^-, \widetilde{v}_2^-, ..., \widetilde{v}_n^-) = \left\{ \min_1 v_{ij} \middle| \begin{matrix} i = 1,2,...,m \\ j = 1,2,...,n \end{matrix} \right\}$$

The distance of each alternative from A+ and A− can be currently calculated using Equations [19] and [20].

$$(20)$$

$$d_i^+ = \sum_{j=1}^n d(\widetilde{v}_{ij}, \widetilde{v}_j^+), i = 1,2,...,m$$

$$(21)$$

$$d_i^- = \sum_{j=1}^n d(\widetilde{v}_{ij}, \widetilde{v}_j^-), 1 = 1,2,...,m$$

If a~ = (a1, a2, a3) and b~= (b1, b2, b3) are two triangular fuzzy numbers, then the vertex method is used to calculate the distance between them and is calculated as:

$$(22)$$

$$d(\widetilde{a}, \widetilde{b}) = \sqrt{[(a_1 - b_1)^2 + (a_2 - b_2)^2 + (a_3 - b_3)^2]}$$

At the end, the relative closeness of each alternative to the ideal solution is calculated as below:

$$(23)$$

$$RC_j = \frac{d_j^-}{d_j^+ + d_j^-} \ \ j = 1,...,m$$

RCi is then used to rank the alternatives. The higher the RCi, the higher criticality of physical server. The higher value of RCi indicates that an alternative is closer to the positive ideal solution and farther from the negative ideal solution simultaneously. A value of 1 (or 100 per cent) for an alternative indicates that the alternative is equal to the positive ideal solution and a value of 0 (or 0 per cent) is equal to the negative ideal solution. The best alternative is the one with the greatest relative closeness to the positive ideal solution.

## 3.2 The fuzzy decision making[4] software tool

The fuzzy decision making (FDM) software tool[55] has been prepared to make decisions onsidering specific criteria and the effect of qualitative parameters and in the situation where the decision maker does not have access to precise information, by Meamareiani at the engineering faculty of Tarbiat-modares University in Iran. The FDM software tool has been designed based on the Fuzzy TOPSIS technique, presented in the previous section. In this section, it is attempted to remove the hotspot in existing cluster nodes by using this software tool and applying it systematically. The most important advantage of applying this software tool is its ability in cases where diversity of data exists.

## 5 A NEW MODEL FOR FINDING MOST LOADED SERVER IN CLUSTER SERVER

So far, the problems with methods to sort physical server have been discussed. Now our technique applying fuzzy decision making software tool in the process of selecting the most overloaded server is presented. The main purpose of this paper is to present a fuzzy multi-criteria decision making model to sort physical nodes from the most to the least loaded.

The basic idea is to move load from the most overloaded servers to the least-overloaded servers, while attempting to minimize data copying incurred during migration. A server can be overloaded along one or more dimensions− CPU, network , memory, temperature and ….if multiple resources of a nodeare heavily utilized, the corresponding serever position is near to the top in the sorting table and results in a correspondingly higher score. The volume captures the degree of (over)load along multiple dimensions in a weighted fashion and can be used by the mitigation algorithms to handle all resource hotspots. This paper presents a hierarchical process based on TOPSIS that operated in two steps. In step one, all physical nodes of the Cluster are being compared to each other. After applying decision algorithm, nodes are sorted fram the highest overloaded to the idle ones (if available). Some parameters used in this stage are shown in Table 1. Every node will be labled with a number between 0 and 1(0 means idle server and 1 indicates saturated server). If there is a node beaking the threshold we triy to mitigate its load by migration the most appropriate Virtual Machine runing on that machine to the least loded node determined in the previous step.

---

[4] FDM

At the next level we come to the most critical node and apply the fuzzy decision algorithm for the second time. The result of this level will be finding the best VM candidate for migration. In this step, we use a new set of parameters that some of them are not exist in the first stage (Table2).

<div align="center">

TABLE1

FIRST LEVEL PARAMETERS

</div>

| N | Name | Data Type | Type | Weight | Description |
|---|------|-----------|------|--------|-------------|
| 1 | CPU% | Deterministic | Benefit | VH | CPU usage |
| 2 | RAM % | Deterministic | Benefit | ML | RAM usage |
| 3 | NET % | Deterministic | Benefit | ML | NET usage |
| 4 | # VM | Lingustic | Benefit | L | Number of VM on Node |
| 5 | CPU cycle | Deterministic | Cost | VH | CPU clock speed(GHZ) |
| 6 | NET BW | Deterministic | Cost | ML | Band width of each Node |
| 7 | TMP | Fuzzy | Benefit | M | Host Temperature |
| 8 | RAM capacity | Deterministic | Cost | ML | RAM of Host(GB) |

For example, in Table I, the variety of the data that can be explained to illustrate the server condition has been offered. In tests, five parameters were used. The first one is the CPU usage of every node in percent.Because of assuming the cluster heterologous,in addition to the CPU usage, the clock speed of the processors is important too. For more explanation, considering two hetrogenous1.8 GHZ and 3.2 GHZ physical nodes with two different CPU usage (fore example 60% and 75%), the fisrt one is more dangerous and occurring saturation is more possible although its utilization hs lower the second node. If comparision made overe the utilization of CPU parameter only, the second server hade been choosed. Therfore, to decide more efficient, both of parameters are necessary.These informations are deterministics and mentioned as percent and numbers. RAM utilization and RAM capacity are another criterias which have incredible role in server load.The next pair of parameters are Network utilization and server Network band width.

## 5.1 TOPSIS methode

There are some methods for solving Multiple Criteria Decision-Making problems, of which one is the TOPSIS

<div align="center">

TABLE2

SECOND LEVEL PARAMETERS

</div>

| N | Name | Data Type | Type | Weight | Description |
|---|------|-----------|------|--------|-------------|
| 1 | CPU% | Deterministic | Benefit | VH | CPU usage by VM |
| 2 | RAM % | Deterministic | Benefit | ML | RAM usage by VM |
| 3 | NET % | Deterministic | Benefit | ML | BW usage by VM |
| 4 | RAM usage | Deterministic | Cost | H | RAM used by VM(GB) |
| 5 | QoS | Linguistic | Benefit | H | Quality of Service for VM |

method.The principle behind TOPSIS is simple: The chosen alternative should be as close to the ideal solution as possible and as far from the negative-ideal solution as possible. The ideal solution is formed as a composite of the best performance values exhibited (in the decision matrix) by any alternative for each attribute. The negative-ideal solution is the composite of the worst performance values. Proximity to each of these performance poles is measured in the Euclidean sense (e.g., square root of the sum of the squared distances along each axis in the "attribute space"), with optional weighting of each ttriute.

## 5.2 PARAMETER TYPE

TOPSIS algorithm can receive three types of information, including deterministic, linguistic, and fuzzy information. These three types of data are indeed parameters affecting the decision making process for selecting the overloaded servers. But in previous methods only the crisp (ordinary) values constitute the decision making process input.

Number of VM in every host is a linguistic parameter. Specification of each node differ from others. It means that the number of virtual machine that can handle is different. For example 4 virtual machines do not impose an annoying load on a specified server while over another weak node available on the cluster can not be handeled.Therefore, we mentione this parameter linguistically. In the FDM software tool, the linguistic variables divided to seven-levels, 'very low', 'low', 'more or less (MoL) low', 'medium', 'more or less (MoL) high', 'high' and 'very high'. To apply mathematical formula ove linguistic statement, we need to map them with numbers. On way is presented by Saaty. His technique is in table5. In this table numbers 9 is assigned to VH and 1 to VL. Other values are between these two numbers.

To change these description to value Based on these assumptions, a transformation table can be created as shown in Table 4.

Physical host temperature is one of the parameters used to make better decision. Although there is a relationship between load of a server and its parameter, sometimes changing in temperature with high amplitude is asigh of failure in the system. If the cooling component is down or works poorly the temperature of the node rising suddenly and as a concequense server may failed.So, this parameter can be used as a predicting failure mechanisim. On the other hand, operating temperature varies from one sever to another depending on the hardware configuration.As a result, it is agood idea to describe this parameter as a Fuzzy statement. Regarding fuzzy numbers related to the temperature of every active node are shown in Table III. it should be added that in our experiments, temperature of every host sremeasure in kelvin per  and a and b values are left and right limit of the main value of temperature. For the Fuzzy parameter, we use Triangular Fuzzy Number. The decision making process in a fuzzy environment is the same as the decision making process in the human brain, because in everyday life, people analyses much inaccurate fuzzy information and then makes decisions.

To more flexibility of our algorithm, we defined a parameter named QoS. It stands for quality of service and indicates the importance of each virtual machine. In the other word more QoS degree for a VM forced the migration control unit to more pay attention to. If  two VM have the same condition, VM with high QoS will be migrated to use more resources on the destination node. Tis parameter is linguistic.

TABLE 3

TRANSFORMATIN FOR FUZZY MEMBERSHIP FUNCTION

| Rank | Abbreviated | Membership function |
|------|-------------|---------------------|
| Very Low | VL | (30,0,10) |
| Low | L | (40,10,10) |
| Mol Low | ML | (50,10,10) |
| Medium | M | (60,10,10) |
| Mol High | MH | (70,10,10) |
| High | H | (80,10,10) |
| Very High | VH | (90,10,0) |

TABLE 4

TRANSFORMATION TABLE

| Rank | Number |
|------|--------|
| Very Low | 1 |
| Low | 3 |
| Mol Low | 4 |
| Medium | 5 |
| Mol High | 6 |
| High | 7 |
| Very High | 9 |

## 6  IMPLEMENTATION AND EVALUATION

To implement our proposed technique, a cluster server with five nodes were applied.Each node has some virtual machines. As Table 5 indicates we have totally 12 virtual machine and for initiation assign RAM to each one. The virtual machines run a mix of  applications ranging from Apache and streaming servers to PHP, and MySQL. We run RUBiS on our servers--RUBiS is an open-source multi-tier web application. When we start running the VMs over five node of our cluster, workload of each server varies over time. To see the result of our migration algorithm, workload of VMs are set to have their peak in predefined time(see Table 5). Fore example PM2 has one virtual machine which is idle so there is no peak in its workload while in PM3 there are four VMs named VM1, VM2, VM3 and VM4 that VM2 is constant load. On the other word, VM1, VM3, and VM4 see their peaks on 450, 400, and 550 second after the cluster starts working respectively. Control unit runs every 3 minutes because some of the workloads change quickly. Depends on the VMs nature this interval can be set. Information from all nodes come to the control unit and verify every three minutes to see every hotspot. After inserting information of nodes in to the algorithm(Fig.1). Decision technique tries to sort servers from high to low score. In order to remove a dimension, the decision matrix is normalized and calculated using weighted normalized ratings automatically.The next action is to find the negative as well as the positive ideal solutions. After finding the ideal and negative solutions, the distance of each alternative is obtained in an n-dimensional space (n is the number of criteria affecting decision making).

<div align="center">

T<small>ABLE</small> 5

N<small>ODES OF CLUSTER UNDER TEST AND CORRESPONDING VIR-</small>
<small>TUAL MACHINES</small>

</div>

| PM | VM | Predefined RAM(MB) | Peak time(s) |
|----|----|----|----|
| PM1 | VM 5 | 256 | 10 |
|  | VM6 | 256 | ----------- |
| PM2 | VM7 | 128 | ----------- |
| PM3 | VM1 | 512 | 450 |
|  | VM2 | 128 | ----------- |
|  | VM3 | 128 | 400 |
|  | VM4 | 256 | 550 |
| PM4 | VM8 | 128 | 100 |
|  | VM9 | 256 | ----------- |
|  | VM10 | 128 | 235 |
| PM5 | VM11 | 128 | 134 |
|  | VM12 | 256 | ----------- |

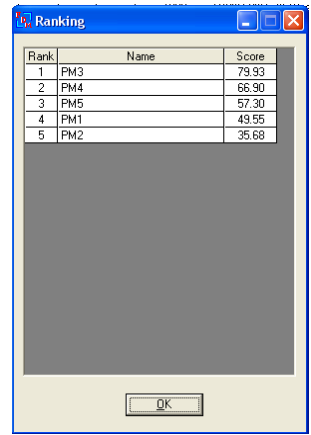

Fig.2.Results of first stage

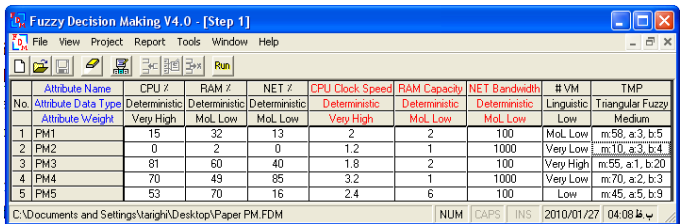

Fig.1.Fuzzy Decision Making(stage 1)

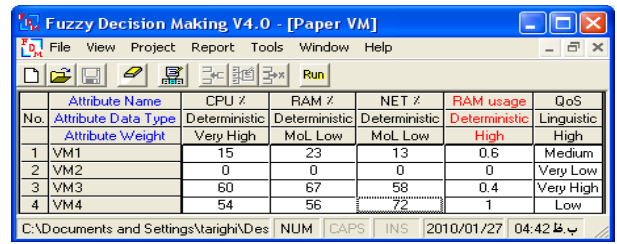

Fig.3.Finding VM to migrate(stage 2)

RAM usage is the parameter that has Cost Type, consequently the TOPSIS algorithm find the VM which has the lowest RAM usage to avoid of transferring larg data over cluster. Result of the second program identified VM3 as the best candidate to migrate(Fig.4).

The final scores of each parameter is its relative closeness to the positive ideal solution. These processes are performed by the algorithm tool and the user enters only the input information such as selection criteria, their effective weight and selection alternatives.As figure 2 shows the physical server 3 has obtained the highest score.Then control unit verifies if this volume braeks the threshold define by the administrator or not.If crossing accures It means that this node is in danger and migration is inevitable. During our test, the threshold set to 75. In example shows in figure 1. Physical node 3 has been overloaded and breaks the threshold. Therefore; the output of this program is a list of server which PM3 is at top and PM2 is the last entry(Fig.2). In next stage another program runs over a different set of parameter to find the VM causing hotspot in PM3(see Fig.3).

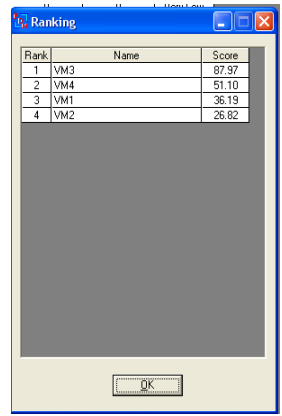

Fig.4.Result of stage 2

If we do not use algorithm to balance the load of the cluster, some nodes will down and services on them stop working. Fore example, node PM3 see bottleneck after 400 seconds. Load of node PM3 has spike at 400, 450, and 550 seconds. Without applying a technique to decrease of load over PM3, this node will be overloaded. Fig.5 indicates this situation. Moreover; Workload of every nodes over time ha been showed there.

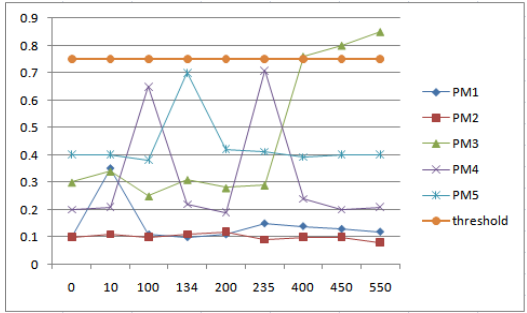

Fig.5.PMs load over time and the threshold

Using load balancing control unit we can detedct hotspot on PM3 and via migrate VM3 from PM3 to PM2 (the least loaded server) the load will be distributed better and response time of the cluster improved(see Fig.6). In this figure, load of PM3 by reloacating VM3 decreased to about 60% and the hot spot was eliminated. On the other hand. Workload on PM2 increased because it hosted VM3.

If sufficient resources are not available, then the algorithm examines the next least loaded server and so on, until a match is found for the candidate VM. If no physical server can house the selected VM, then the algorithm moves on to the next VM and attempts to move it in a similar fashion. The process repeats until the utilizations of all resources on the physical server fall below the thresholds.

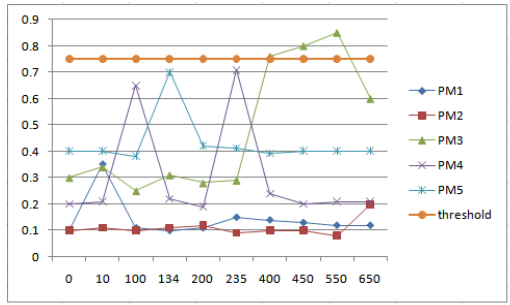

Fig.6.Eliminating hot spot of node PM3

## 8 ALGORITHM OVERHEAD

Although the disadvantages of numerical ranking methods have been removed partly in the decision making models offered, this methods has its own limits. In decision making models, which are based on multicriteria decision making techniques, there is no limitation on the number of criteria and alternatives, but these models face the problem of time-consuming calculations(Fig.7). This figure illustrates that complexity of TOPSIS algorithm increases whe the number of VMs run on the cluster increse. Reducing the migration overhead (i.e., the amount of data transferred) is important, since Xen's live migration mechanism works by iteratively copying the memory image of the VM to the destination while keeping track of which pages are being dirtied and need to be resent. This requires Xen to intercept all memory accesses for the migrating domain, which significantly impacts the performance of the application inside the VM. By reducing the amount of data copied over the network, we can minimize the total migration time, and thus, the performance impact on applications. Note that network bandwidth available for application use is also reduced due to the background copying during migrations. In sandpiper[67], to determine which VMs to migrate, the algorithm orders physical servers in decreasing order of their volumes. Within each server, VMs are considered in decreasing order of their volume-to-size ratio (VSR); where VSR is defined as Volume/Size; size is the memory footprint of the VM. By considering VMs in VSR order, the algorithm attempts to migrate the maximum load per unit byte moved, which has been shown to minimize migration overhead [20]. In our algorithm, to minimize the migrating data, another property of TOPSIS technique has been applied.For more illustration, with referring to Table1,we can see that there is a column in the table that labled with Type. This column gets two word, Benefit or Cost. In the other word, when we lable a parameter as Cost,it means that we want to select the Alternative which has this parameter as less possible as. For the Benefit, it is vise versa. As shown in Figure 3, Type of parameter RAM usage is cost because we want to choose a VM from all virtual machines run simultaneously on PM3 that has more CPU%,RAM %, NET %,QoS quantity but less RAM usage.

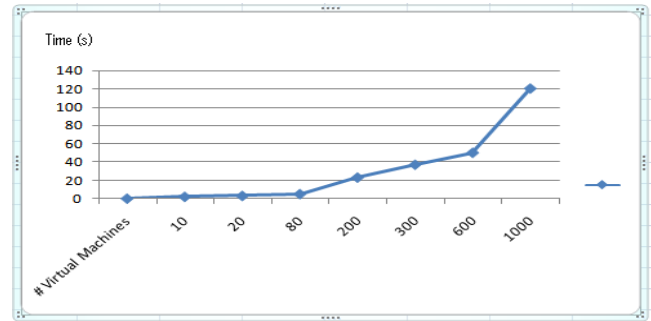

Fig.7.Time used to run TOPSIS algorithm when number of virtual machines increase

## 9 CONCLUSIONS AND FUTURE WORK

The hot node sorting of virtualized cluster servers is a critical point and strategic issue in the migratin process. This decision involves many parameters that are interrelated in that changes in some parameters affect the others.

This paper has discussed cluster node selection in a fuzzy environment and uncertain linguistic value of variables. Fuzzy TOPSIS is a viable method for the proposed problem and is suitable for the use of linguistic variables. When the decision making condition is vague and inaccurate, then this method is the preferred technique. The present study explored the use of Fuzzy TOPSIS in finding the most critical server and the least one.

The proposed model can be a suitable tool to rank servers. A real case was studied and Fyzzy selection algorithm applied.

The systematic evaluation by Fuzzy TOPSIS of machine selection problems can reduce the risk of a poor choice. The Fuzzy TOPSIS is one of the compensatory decision making methods. As mentioned before, in this method decreasing the score of one parameter is compensated by increasing the score of other parameter(s) and vice versa. Moreover, there is no limitation on the number of alternatives and criteria. By applying the FDM model, based on Fuzzy TOPSIS, a strategy was offered to extract a list of server from the most to the least loaded. This strategy has advantages in comparison with previous numerical ranking (scoring) methods such as Sandpiper. These advantages are a strong theoretical base on fuzzy logic, the ability of sensitivity analysis, the direct usage of linguistic variables in the selection process, unlimited alternatives and criteria, and, most important of all, the possibility of considering the mutual effects of different parameters in the selection process. In fact, TOPSIS is one of the compensatory multiattribute decision making models. Moreover, this model considers the uncertainty associated with the input parameters (linguistic variables) used in the selection process.

## ACKNOWLEDGMENT

The authors wish to thank Mis Zahra Palizdar for her contribution in developing this article.

**M.Tarighi** is now PHd student in Tehrn Polytechnique University. He graduated from SHAHID SHAMRAN university of AHWAZ in BSc of Electronics & from Amir Kabir University of technology in MSc of Electronics. His interests are currently Cluster Computing, Virtualization, and Decision Algorithms.Beside that, soccer, Internet, Volunteer works, chess, cross word puzzle, and poetry could be mentioned.